\documentclass[prb,a4paper,showpacs,floatfix,superscriptaddress,secnumarabic,twocolumn]{revtex4}
\usepackage[english]{babel}
\usepackage{amsmath,amssymb}
\usepackage{dsfont}
\usepackage{txfonts}
\usepackage{upgreek}
\usepackage{subfigure}
\usepackage{graphicx}
\usepackage{dcolumn}
\usepackage{float}
\usepackage{ifpdf}
\ifpdf
  \usepackage[dvipdfm,colorlinks,hyperindex]{hyperref}
\else
  \usepackage[hypertex,colorlinks,hyperindex]{hyperref}
\fi

\makeatletter

\newcommand{\mces}{\mathcal{E}_{so}}
\newcommand{\mceg}{\mathcal{E}_{g}}
\newcommand{\mcee}{\mathcal{E}_{e}}
\newcommand{\mceo}{\mathcal{E}_{o}}
\newcommand{\mce}{\mathcal{E}}
\newcommand{\varF}{\mathcal{F}}
\newcommand{\varH}{\mathcal{H}}

\newcommand{\bfA}{\mathbf{A}}
\newcommand{\bfp}{\mathbf{p}}
\newcommand{\bfz}{\mathbf{z}}
\newcommand{\bfnabla}{{\boldsymbol{\nabla}}}
\newcommand{\bfPi}{{\boldsymbol{\uppi}}}

\newcommand{\sgn}{\mathop{\operator@font sgn}}

\newcommand{\ket}[1]{|#1\rangle}
\newcommand{\Braket}[1]{\mathinner{\langle{\textstyle#1}\rangle}}
\newcommand{\ode}[3][]{\frac{d^{#1}{#2}}{d{#3}^{#1}}}
\newcommand{\set}[1]{\left\{#1\right\}}

\newcommand{\uz}{\hat{\bfz}}

\newcommand{\eqnref}[1]{Eq.~(\ref{#1})}
\newcommand{\eqnsref}[1]{Eqs.~(\ref{#1})}
\newcommand{\figref}[1]{Fig.~\ref{#1}}
\newcommand{\Figref}[1]{Figure~\ref{#1}}

\newcommand{\up}{\uparrow}
\newcommand{\down}{\downarrow}

\newcommand{\imsizeA}{0.475\columnwidth}
\newcommand{\imsizeAb}{0.41\columnwidth}
\newcommand{\imsizeB}{0.950\columnwidth}

\makeatother

\begin{document}
\title{Spin Hall effect due to intersubband-induced spin-orbit interaction in symmetric quantum wells}
\author{Minchul Lee}
\affiliation{Department of Applied Physics, Kyung Hee University, Yongin 449-701, Korea}
\author{Marco O. Hachiya}
\affiliation{Instituto de F\'{\i}sica de S\~ao Carlos, Universidade de S\~ao Paulo, 13560-970 S\~ao Carlos, S\~ao Paulo, Brazil}
\author{E. Bernardes}
\affiliation{Instituto de F\'{\i}sica de S\~ao Carlos, Universidade de S\~ao Paulo, 13560-970 S\~ao Carlos, S\~ao Paulo, Brazil}
\author{J. Carlos Egues}
\affiliation{Instituto de F\'{\i}sica de S\~ao Carlos, Universidade de S\~ao Paulo, 13560-970 S\~ao Carlos, S\~ao Paulo, Brazil}
\affiliation{Department of Physics, University of Basel, CH-4056 Basel, Switzerland}
\author{Daniel Loss}
\affiliation{Department of Physics, University of Basel, CH-4056 Basel, Switzerland}

\begin{abstract}
  We investigate the intrinsic spin Hall effect in two-dimensional electron
  gases in quantum wells with two subbands, where a new intersubband-induced
  spin-orbit coupling is operative. The bulk spin Hall conductivity
  $\sigma^z_{xy}$ is calculated in the ballistic limit within the standard Kubo
  formalism in the presence of a magnetic field $B$ and is found to remain
  finite in the $B=0$ limit, as long as only the lowest subband is
  occupied. Our calculated $\sigma^z_{xy}$ exhibits a non-monotonic behavior
  and can change its sign as the Fermi energy (the carrier areal density
  $n_{2D}$) is varied between the subband edges. We determine the magnitude of
  $\sigma^z_{xy}$ for realistic InSb quantum wells by performing a
  self-consistent calculation of the intersubband-induced spin-orbit coupling.
\end{abstract}
\date{\today}

\pacs{
  72.25.Dc, 
  73.21.Fg, 
  71.70.Ej  
}

\maketitle

\section{Introduction}

The Spin Hall effect (SHE) \cite{Engel07} refers to the spin accumulation with
opposite polarizations at the two edges of a Hall bar, due to the transverse
spin current induced by a driving longitudinal charge current. This effect was
first theoretically proposed \cite{ESHE} as arising from the spin-dependent
scattering at impurities with spin-orbit interaction. In the literature this is
commonly referred to as the ``extrinsic SHE'' as it relies on the presence of
impurities (``extrinsic mechanism''). In this case, the asymmetric Mott-skew
and side jump scattering contributions drive spin up and down electrons toward
opposite directions, thus giving rise to a net transverse spin current (with no
charge current) and to spin accumulation at the edges.  Recently, two
theoretical works\cite{Murakami03,Sinova04} predicted that the spin-orbit
effects on the band structure of semiconductors -- the so called ``intrinsic
mechanism'' for the SHE -- can also give rise to a spin current perpendicular
to an applied electric field, even in the absence of impurities. These authors
have calculated the ballistic spin Hall conductivity (SHC) $\sigma^z_{xy}$,
defined via $J^z_x = \hbar j^z_x /2 = \hbar (v_x \sigma_z + \sigma_z v_x)/4 =
\sigma^z_{xy} E_y$, where $v_x$ is the $x$ component of the velocity operator
and $\sigma_{x,y,z}$ are the Pauli matrices, for a p-doped three-dimensional
(bulk) valence band system \cite{Murakami03} and for a two-dimensional electron
gas (2DEG) with the Rashba spin orbit interaction.\cite{Sinova04}

A number of theoretical papers have investigated the robustness of the
ballistic SHC $\sigma^z_{xy}$ as arising from the intrinsic mechanism, against
scattering by
non-magnetic\cite{Inoue04,Mishchenko04,raimondi05,Chalaev05,Dimitrova05,Malshukov05,Shen}
and magnetic\cite{Gorini08} impurities, its dependence on specific classes of
SO interactions,\cite{Malshukov05} and the interplay between the SO coupling
and magnetic fields.\cite{Shen,Rashba04,Lucignano08} Two early experimental
efforts \cite{Kato04,Wunderlich05} have probed the spin Hall effect for
electrons and holes.\cite{Schliemann05} Kato \textit{et al.}\cite{Kato04} find
that the SHE in n-GaAs epilayers is due to the extrinsic mechanism
(non-magnetic impurities), while Wunderlich \textit{et al.}\cite{Wunderlich05}
conclude that the SHE in the 2D hole gas probed in their experiment is
intrinsic. Further investigations have also shown that the SHE in 2DEGs is of
the extrinsic type.\cite{Sih05}

Following an early debate concerning the robustness of the intrinsic SHE, it is
now well established that the dc SHC as defined above vanishes identically for
model Hamiltonians with a linear-in-the-carrier-momentum SO interaction, such
as that of Rashba and/or the linearized Dresselhaus. This holds in both the
ballistic case\cite{Rashba04} (``clean limit'') and in the limit of weak
scattering by non-magnetic
impurities.\cite{Inoue04,Mishchenko04,raimondi05,Chalaev05,Dimitrova05,Malshukov05,Shen,Sinova06}
This result can be understood by examining the relationship between the time
derivative of the spin density and spin current in these systems. As pointed
out in Refs.~[\onlinecite{Erlingsson05}], [\onlinecite{Chalaev05}], and
[\onlinecite{Dimitrova05}], for Rashba-type models $d\sigma_k/dt \propto
j^z_k$, $k=x,y$ and should vanish in the dc steady state regime (in the
presence of some relaxation mechanism) where
$d\sigma_k/dt=0$.\cite{non-zero-she}

The search for new materials that can exhibit the SHE as well as other types of
SO interactions has continued over the years.\cite{Koenig09} Recently, we have
introduced a new type of SO interaction present in III-V (or II-VI) zinc-blende
semiconductor quantum wells with more than one
subband.\cite{Bernardes06,Calsaverini08} This intersubband-induced SO term is
similar in form to the Rashba SO interaction. However, it couples electron
states from distinct subbands and hence can be non-zero even in structurally
symmetric wells.  For an electron in a symmetric quantum well with two subbands
we have\cite{Bernardes06,Calsaverini08}
\begin{equation}
  \label{eq:H}
  \varH
  =
  \left(\frac{\bfp^2}{2m} + \bar\mce\right) \mathds{1}\otimes\mathds{1}
  - \frac{\Delta\mce}{2}\, \tau_z\otimes\mathds{1}
  + \frac{\eta}{\hbar} \tau_x\otimes (p_x\sigma_y - p_y\sigma_x),
\end{equation}
where $m$ is the effective mass, $\bar\mce = (\mcee+ \mceo)/2$ and $\Delta\mce
= \mceo-\mcee$, with $\mcee$ and $\mceo$ denoting the band edges of the lowest
(even) and first excited (odd) subbands. $\tau_{x,y,z}$ and $\sigma_{x,y,z}$
are Pauli matrices describing the subband degree of freedom and the electron
spin, respectively. The intersubband-induced SO coupling $\eta$ depends on the
structural potential of the well, the electronic Hartree potential, and the
external gate potential.\cite{Calsaverini08}

In this paper we calculate the dc spin Hall conductivity $\sigma^z_{xy}$ for 2D
electrons in the presence of the intersubband-induced SO interaction in wells
with two subbands. We use the Kubo formula in the ballistic limit.  We follow
Rashba's approach\cite{Rashba04} by performing our calculation in the presence
of a perpendicular magnetic field $B$, which modifies the energy spectrum
(Landau levels) thus allowing us to consistently include intra- and
inter-branch transitions in the Kubo formula, and then taking the $B\to0$
limit. With this procedure, known to produce the correct vanishing of
$\sigma^z_{xy}$ for the Rashba model,\cite{Rashba04} we derive an analytical
expression for the dc SHC in two subband systems. More specifically, we find
that: (i) $\sigma^z_{xy}$ is non-zero for electrons whose Fermi energy $\mce_F$
lies between the subband edges ${\mcee}$ and ${\mceo}$, i.e., when only the
lowest subband is occupied, and (ii) $\sigma^z_{xy}$ is null for
$\mce_F>\mceo$, i.e., when two subbands are occupied. Interestingly, the
non-zero SHC is non-universal (e.g., it depends on $\eta$), exhibits a
non-monotonic behavior and a sign change for $\mcee<\mce_F<\mceo$. In addition,
the SHC presents finite jumps at the subband edges when plotted as a function
of $\mce_F$ (or the carrier areal density $n_{2D}$), due to discontinuities in
the density of states contributing to the spin density. We have also performed
a detailed self-consistent calculation of the energy spectrum and wave
functions for realistic wells, from which we determine $\eta$ and the
corresponding $\sigma^z_{xy}$. The magnitude of our self-consistently
determined SHC is, however, much smaller than $\sigma_0=e/4\pi$.\cite{spin-qc}

Before describing in detail our linear response calculation for the SHC, here
we present a simple argument as to why we could expect a non-vanishing SHC for
electrons in 2DEGs with two subbands where the intersubband SO interaction is
operative. Despite the formal similarity between the SO term in our
Hamiltonian, \eqnref{eq:H} and that of Rashba, here we find a different
relationship between the spin current and spin density, i.e.,
\begin{align}
  \label{eq:jzx}
  j^z_x & = \frac{\hbar^2}{2m\eta}
  \left(
    \ode{}{t}(\tau_x\otimes\sigma_x)
    -
    \frac{\Delta\mce}{\hbar}\tau_y\otimes\sigma_x
  \right),
\end{align}
in contrast to the Rashba model for which $d\sigma_x/dt \propto j^z_x$.  In
\eqnref{eq:jzx} the spin current is related to the subband-related spin
densities $\tau_x\otimes\sigma_x$ and $\tau_y\otimes\sigma_x$. The second term
in \eqnref{eq:jzx}, absent in the Rashba system, suggests that the (pseudo)
spin density response to an applied electric field can contribute to the spin
current -- even in the dc steady state limit where
$d(\tau_x\otimes\sigma_x)/dt=0$. Hence, 2DEGs formed in two subband wells with
intersubband SO coupling may have a non-zero SHC. Our detailed linear response
calculation below shows that this is indeed the case in the clean limit.

\section{Model Hamiltonian and Kubo Formula}

Let us consider our Hamiltonian \eqnref{eq:H} in the presence of a magnetic
field $B\uz = \bfnabla\times\bfA$ by making the replacement $\bfp \to \bfPi =
\bfp - (e/c)\bfA$. For simplicity, we do not consider the Zeeman splitting
(i.e, we assume a zero $g$ factor\cite{Lucignano08}). Since our SO term couples
only electrons with opposite spins in different subbands, the subband-spin
Hilbert space $\set{\ket{bs}; b={e,o}, s=\up,\down}$ can be divided into two
independent subspaces $\varF_+ = \set{\ket{{e}\up},\ket{{o}\down}}$ and
$\varF_- = \set{\ket{{o}\up},\ket{{e}\down}}$, and the $2\times2$ Hamiltonian
in each subspace $\lambda=\pm$ can be written as
\begin{equation}
  \varH_\lambda
  =
  \frac{\bfPi^2}{2m}
  +
  \bar\mce
  -
  \begin{bmatrix}
    \lambda \Delta\mce/2 & \eta (\uppi_y {+} i \uppi_x)/\hbar
    \\
    \eta (\uppi_y {-} i \uppi_x)/\hbar & - \lambda \Delta\mce/2
  \end{bmatrix}.
\end{equation}
The reduced Hamiltonian is identical to that of a 2DEG with a Rashba SO
coupling of strength $\eta$ and an effective Zeeman splitting $\lambda
\Delta\mce$. Note that only the sign of the effective Zeeman splitting differs
between $\varH_\lambda$ in two subspaces.

In the absence of the SO coupling $(\eta=0)$, the Schr\"odinger equation gives
rise to the Landau levels $\ket{bsn}_0$ with the level index $n\ge0$ for each
subband $b$ and spin $s$. The Landau levels are evenly spaced by the cyclotron
gap $\hbar\omega_c=|eB|/mc$.  The degeneracy (per unit volume) of each level
and the magnetic length are, respectively, $1/2\pi l^2$ and $l =
\sqrt{c\hbar/eB}$. For non-zero SO coupling $(\eta\ne0)$ the coupling of Landau
levels within the same subspace produces the mixed Landau levels
$\ket{\lambda\mu n}$:
\begin{equation}
  \begin{bmatrix}
    \ket{\lambda{+}n}
    \\
    \ket{\lambda{-}n}
  \end{bmatrix}
  =
  \begin{bmatrix}
    \displaystyle \sin\frac{\theta_{\lambda n}}{2}
    & \displaystyle - i\cos\frac{\theta_{\lambda n}}{2}
    \\
    \displaystyle \cos\frac{\theta_{\lambda n}}{2}
    & \displaystyle - i\sin\frac{\theta_{\lambda n}}{2}
  \end{bmatrix}
  \begin{bmatrix}
    \ket{b_1{\up}n}_0
    \\
    \ket{b_2{\down}n{-}1}_0
  \end{bmatrix},
\end{equation}
where $(b_1,b_2) = ({e,o})$ for $\lambda{=}+$ and $({o,e})$
for $\lambda{=}-$, with the corresponding eigenenergies
\begin{equation}
  \mce_{\lambda\mu n}
  =
  \bar\mce
  +
  \hbar\omega_{\rm c} \left(n - \mu\gamma_{\lambda n}\right).
\end{equation}
Here $\mu = \pm$ denotes the spin branch. It is convenient to introduce
dimensionless parameters $\mces = 2m\eta^2/\hbar^3\omega_{\rm c}$ and $\mceg =
\Delta\mce/\hbar\omega_{\rm c}$ for the SO coupling energy and the subband gap,
respectively. In term of these we can define $\sin\theta_{\lambda n} =
\sqrt{n\mces}/\gamma_{\lambda n}$ and $\cos\theta_{\lambda n} =
\zeta_\lambda/\gamma_{\lambda n}$ with $\zeta_\lambda = (1 - \lambda\mceg)/2$
and $\gamma_{\lambda n} = \sqrt{n\mces + \zeta_\lambda^2}$.  Note that the
eigenstate for $n=0$ exists only for $\mu = - \sgn\zeta_\lambda$.

We determine the dc spin Hall conductivity at zero temperature by using
the Kubo formula \cite{Rashba04}
\begin{equation}
  \label{eq:SHC}
  \sigma^z_{xy}
  =
  - \frac{ie}{2\pi l^2} \sum_\lambda\!\!
  \sum'_{\mu n\mu'n'}
  \frac{\Braket{\lambda\mu n|j^z_x|\lambda\mu'n'}
    \Braket{\lambda\mu'n'|v_y|\lambda\mu n}}%
  {(\omega_{\lambda\mu n} - \omega_{\lambda\mu'n'})^2},
\end{equation}
where the primed sum indicates that it should be performed over the states with
$\omega_{\lambda\mu n} = \mce_{\lambda\mu n}/\hbar < \mce_F/\hbar$ and
$\omega_{\lambda\mu'n'} > \mce_F/\hbar$. Here we have used the fact that the
operators $j^z_x$ and $v_y$ couple only states in same subspace and the
matrices $\Braket{\lambda\mu n|j^z_x|\lambda\mu'n'}$ and
$\Braket{\lambda\mu'n'|v_y|\lambda\mu n}$ are symmetric and antisymmetric,
respectively. Since the matrix elements are independent of the guiding center
position, the factor $1/2\pi l^2$ appears due to the Landau level degeneracy.

A more insightful analysis of the SHC can be achieved by expressing the
operators in terms of commutators with the Hamiltonian.  First, the relation
$v_y = (i/\hbar)[\varH,y]$ leads to
\begin{equation}
  \label{eq:vyc}
  \Braket{\lambda\mu'n'|v_y|\lambda\mu n}
  =
  il^2 (\omega_{\lambda\mu'n'} {-} \omega_{\lambda\mu n})
  \Braket{\lambda\mu'n'|k_x|\lambda\mu n},
\end{equation}
with $k_x = (a^\dag + a)/\sqrt2\,l$. This is a direct consequence of the
definition of the ladder operator $a = (l/\sqrt2\,\hbar)(\uppi_x +
i\uppi_y) = (y + l^2 p_x/\hbar + il^2 p_y/\hbar)/\sqrt2\,l$. From
\eqnref{eq:jzx}, on the other hand, we can obtain
\begin{multline}
  \label{eq:jzyc}
  \Braket{\lambda\mu n|j^z_x|\lambda\mu'n'}
  =
  - \frac{\hbar\Delta\mce}{2m\eta}
  \Braket{\lambda\mu n|\tau_y\otimes\sigma_x|\lambda\mu'n'}
  \\
  +
  \frac{i\hbar^2}{2m\eta} (\omega_{\lambda\mu n} {-} \omega_{\lambda\mu'n'})
  \Braket{\lambda\mu n|\tau_x\otimes\sigma_x|\lambda\mu'n'}.
\end{multline}
Substitution of \eqnsref{eq:vyc} and \eqref{eq:jzyc} into
\eqnref{eq:SHC} gives rise to $\sigma^z_{xy} = \sigma^{(1)} +
\sigma^{(2)}$, with
\begin{subequations}
  \begin{align}
    \label{eq:shc1}
    \frac{\sigma^{(1)}}{\sigma_0}
    & =
    \sum_\lambda\!\!
    \sum'_{\mu n\mu'n'} \frac{\hbar^2}{im\eta}
    \Braket{\lambda\mu n|\tau_x\otimes\sigma_x|\lambda\mu'n'}
    \Braket{\lambda\mu'n'|k_x|\lambda\mu n},
    \\
    \label{eq:shc2}
    \frac{\sigma^{(2)}}{\sigma_0}
    & =
    \sum_\lambda\!\!
    \sum'_{\mu n\mu'n'} \frac{\hbar\Delta\mce}{m\eta}
    \frac{\Braket{\lambda\mu n|\tau_y\otimes\sigma_x|\lambda\mu'n'}
      \Braket{\lambda\mu'n'|k_x|\lambda\mu n}}%
    {\omega_{\lambda\mu n} - \omega_{\lambda\mu'n'}},
  \end{align}
\end{subequations}
with $\sigma_0 \equiv e/4\pi$.\cite{spin-qc}  Each of the two terms can be
further divided into two contributions coming from two different transitions
classified by the restriction on the spin branches of states to be summed over:
interbranch and intrabranch transitions.\cite{Rashba04}

\begin{figure}[!t]
  \centering
  \subfigure[\label{fig:transA} $\mce_F > {\mceo}$.]
  {\resizebox{\imsizeA}{!}{\includegraphics{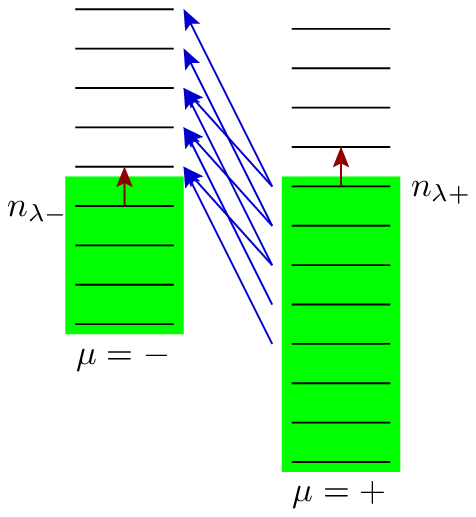}}}\qquad%
  \subfigure[\label{fig:transB} ${\mcee} < \mce_F < {\mceo}$.]
  {\resizebox{\imsizeAb}{!}{\includegraphics{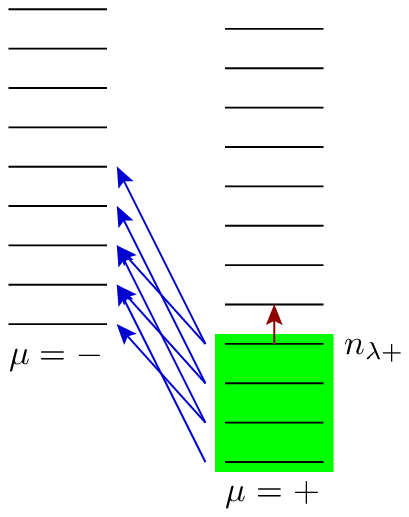}}}
  \par\vspace{-0.25cm}
  \caption{Schematic diagrams describing the interbranch (long arrows) and
    intrabranch (short arrows) transitions. Left (a) and right (b) panels
    correspond to the cases $\mce_F > {\mceo}$ and ${\mcee} < \mce_F <
    {\mceo}$, respectively. $n_\pm$ denotes the highest filled Landau level in
    the spin branch $\mu$.}
  \label{fig:trans}
\end{figure}

As can be seen from \figref{fig:trans}, the interbranch contributions describe
transitions from the filled states in the lower branch $\mu=+$ to the empty
ones in the upper branch $\mu=-$ and follow the selection rule
$\ket{\lambda{+}n} \to \ket{\lambda{-}n{\pm}1}$. The intrabranch transitions
are possible only in the vicinity of the Fermi level in each spin branch and
obey a similar selection rule $\ket{\lambda\mu n_{\lambda\mu}} \to
\ket{\lambda\mu n_{\lambda\mu}{+}1}$, where $n_{\lambda\mu}$ is the index of
the highest filled Landau level in the subspace $\lambda$ and spin branch
$\mu$; see \figref{fig:trans}. In what follows, we write $\sigma^{(i)} =
\sigma^{(i)}_{\rm inter} + \sigma^{(i)}_{\rm intra}$ with $i=1,2$ to identify
and separately investigate the contributions from the interbranch and
intrabranch transitions.

\section{Spin Hall Conductivity}

First we investigate $\sigma^{(1)}$. Making use of the fact that the matrices
$\Braket{\lambda\mu n|\tau_x\otimes\sigma_x|\lambda\mu'n'}$ and
$\Braket{\lambda\mu'n'|k_x|\lambda\mu n}$ are antisymmetric and symmetric,
respectively, one can derive the two following identities
\begin{subequations}
  \label{eq:idn}
  \begin{align}
    \sum_{n\mu n'\mu'}^*
    \Braket{\lambda\mu n|\tau_x\otimes\sigma_x|\lambda\mu'n'}
    \Braket{\lambda\mu'n'|k_x|\lambda\mu n}
    & = 0,
    \\
    \sum_{n'\mu'}
    \Braket{\lambda\mu n|\tau_x\otimes\sigma_x|\lambda\mu'n'}
    \Braket{\lambda\mu'n'|k_x|\lambda\mu n}
    & = 0,
  \end{align}
\end{subequations}
where the asterisk over the sum indicates that it should be done over the
states with $\mce_{n\mu}, \mce_{n'\mu'} < \mce_F$. We have used the commutation
relation $[\tau_x\otimes\sigma_x,k_x] = 0$ to prove the second identity, which
is valid for arbitrary $\lambda$, $\mu$, and $n$. Applying these identities to
\eqnref{eq:shc1} reveals that $\sigma^{(1)}$ vanishes identically for arbitrary
$B$ and $\mce_F$.  Interestingly, the interbranch and intrabranch contributions
exactly cancel out, i.e., $\sigma^{(1)}_{\rm inter} = - \sigma^{(1)}_{\rm
  intra}$. This cancelation is similar to that which occurs in the Rashba
model.\cite{Rashba04} Here, however, the extended operator $\tau_x
\otimes\sigma_x$ replaces the spin operator $\sigma_x$ in the Rashba case. More
explicitly, we have
\begin{equation}
  \label{eq:c1}
  \frac{\sigma^{(1)}_{\rm inter}}{\sigma_0}
  = -\frac{\sigma^{(1)}_{\rm intra}}{\sigma_0}
  = \frac12\sum_\lambda
  \begin{cases}
    \displaystyle
    \frac{n_{\lambda+}}{\gamma_{n_{\lambda+}}}
    -
    \frac{n_{\lambda-}{+}1}{\gamma_{n_{\lambda-}{+}1}},
    & \mce_F {>} {\mceo},
    \\
    \displaystyle
    \frac{n_{\lambda+}}{\gamma_{n_{\lambda+}}},
    & {\mcee} {<} \mce_F {<} {\mceo}.
  \end{cases}
\end{equation}
In the $B\to 0$ limit, we obtain
\begin{equation}
  \frac{\sigma^{(1)}_{\rm inter}}{\sigma_0}
  = -\frac{\sigma^{(1)}_{\rm intra}}{\sigma_0}
  =
  \begin{cases}
    \displaystyle
    \frac{\kappa_1\kappa_2 + 1/2}{\kappa_1\kappa_2 + 1/4},
    & \mce_F {>} {\mceo},
    \\
    \displaystyle
    1 + \frac{\kappa_2}{\kappa_3 + \kappa_1/2},
    & {\mcee} {<} \mce_F {<} {\mceo}
  \end{cases}
\end{equation}
with $\kappa_1 = \mces/\mceg$, $\kappa_2 = (\mce_F - \bar\mce)/\Delta\mce$, and
$\kappa_3 = \sqrt{\kappa_1^2/4 + \kappa_1\kappa_2 + 1/4}$. It is worth noting
that $\sigma^{(1)}_{\rm intra}$ (or $\sigma^{(1)}_{\rm inter}$) varies
continuously as $\mce_F$ passes through the upper subband energy edge
${\mceo}$, even though the intrabranch transitions coming from the $\mu=-$ spin
branch [see \figref{fig:trans}(b)] stop contributing at this energy. This is so
because the contribution $(n_{\lambda-}{+}1)/\gamma_{n_{\lambda-}{+}1}$ in
\eqnref{eq:c1} vanishes as $\mce_F \to \mceo$.

We now evaluate the second term $\sigma^{(2)}$ of the spin Hall conductivity,
\eqnref{eq:shc2}. In our model, $\sigma^{(2)}$ arises solely from the pseudo
spin density response [see \eqnref{eq:jzx}], which has no counterpart in the
Rashba model. Differently than $\sigma^{(1)}$, the expression for
$\sigma^{(2)}$ contains a factor $\omega_{\lambda\mu n} -
\omega_{\lambda\mu'n'}$ in the denominator [cf, \eqnsref{eq:shc1} and
\eqref{eq:shc2}]; this prevents us from deriving identities such as
\eqnref{eq:idn} for $\sigma^{(2)}$.  Hence, no exact cancelation between the
interbranch and intrabranch contributions is guaranteed for $\sigma^{(2)}$. In
general, $\sigma^{(2)}$ is nonzero for arbitrary $B$. In addition, the
denominator factor in $\sigma^{(2)}$ allows for the possibility of a finite
contribution to $\sigma^{(2)}$ even if the matrix elements in the numerator of
\eqnref{eq:shc2} vanish, provided that $(\omega_{\lambda\mu n} -
\omega_{\lambda\mu'n'})$ goes to zero as well. This, in contrast to the
$\sigma^{(1)}$ case, makes the intrabranch contribution $\sigma^{(2)}_{\rm
  intra}$ discontinuous as $\mce_F$ crosses the subband edges.

Explicitly, the interbranch and intrabranch contributions to $\sigma^{(2)}$ are
\begin{widetext}
  \begin{subequations}
    \label{eq:shc2_fB}
    \begin{align}
      \frac{\sigma^{(2)}_{\rm inter}}{\sigma_0}
      & =
      \sum_\lambda \frac{\lambda}{2} \mceg
      \sum_{n=\max(n_{\lambda-}{+}1,0)}^{n_{\lambda+}-1}
      \frac{\displaystyle
        \frac{n+1+\zeta_\lambda}{\gamma_{\lambda n{+}1}}
        - \frac{n-\zeta_\lambda}{\gamma_{\lambda n}}}%
      {(\gamma_{\lambda n{+}1} + \gamma_{\lambda n})^2 - 1},
      \\
      \label{eq:shc2_intra}
      \frac{\sigma^{(2)}_{\rm intra}}{\sigma_0}
      & =
      \sum_\lambda \frac{\lambda}{2} \mceg
      \begin{cases}
        \frac{\displaystyle
          2n_{\lambda+} + 1
          + \frac{n_{\lambda+}-\zeta_\lambda}{\gamma_{\lambda n_{\lambda+}}}}
        {\displaystyle
          1 - \mces + 2\gamma_{\lambda n_{\lambda+}}} -
        \frac{\displaystyle
          2n_{\lambda-} + 1
          + \frac{n_{\lambda-}+1+\zeta_\lambda}{\gamma_{\lambda
          n_{\lambda-}{+}1}}}
        {\displaystyle
          1 + \mces + 2\gamma_{\lambda n_{\lambda-}{+}1}},
        & \mce_F {>} {\mceo},
        \\
        \frac{\displaystyle
          2n_{\lambda+} + 1
          + \frac{n_{\lambda+}-\zeta_\lambda}{\gamma_{\lambda n_{\lambda+}}}}
        {\displaystyle
          1 - \mces + 2\gamma_{\lambda n_{\lambda+}}},
        & {\mcee} {<} \mce_F {<} {\mceo}.
      \end{cases}
    \end{align}
  \end{subequations}
\end{widetext}
The intrabranch contribution from the $\mu=-$ spin branch, i.e., the second
term in \eqnref{eq:shc2_intra}, present only for $\mce_F > {\mceo}$, does not
vanish as $\mce_F \to {\mceo}$ and $B \to 0$. This leads to a discontinuity in
$\sigma^{(2)}$ at $\mce_F = {\mceo}$. A similar discontinuity occurs at $\mce_F
= {\mcee}$ due to the first term in \eqnref{eq:shc2_intra} not vanishing as
$\mce_F \to {\mcee}$ and $B \to 0$.  In the $B\to0$ limit, \eqnref{eq:shc2_fB}
simplifies to
\begin{subequations}
  \label{eq:sig2}
  \begin{align}
    \frac{\sigma^{(2)}_{\rm inter}}{\sigma_0}
    & =
    \begin{cases}
      \displaystyle
      - \frac{1}{2\kappa_1\kappa_2 {+} 1/2}, & \mce_F {>} {\mceo},
      \\
      \displaystyle
      \frac{1}{\kappa_1}\left(\frac{1}{2\kappa_3 {+} \kappa_1} - 1\right),
      & {\mcee} {<} \mce_F {<} {\mceo},
    \end{cases}
    \\
    \frac{\sigma^{(2)}_{\rm intra}}{\sigma_0}
    & =
    \begin{cases}
      \displaystyle
      \frac{1}{2\kappa_1\kappa_2 {+} 1/2}, & \mce_F {>} {\mceo},
      \\
      \displaystyle
      \frac{1}{2\kappa_3 (2\kappa_3 {+} \kappa_1)}
      + \frac{\kappa_2 {+} \kappa_1/2}{4\kappa_3^3},
      & {\mcee} {<} \mce_F {<} {\mceo}.
    \end{cases}
  \end{align}
\end{subequations}

Interestingly, the above results show that the interbranch and intrabranch
contributions to $\sigma^{(2)}$ exactly cancel out in the $B\to0$ limit,
provided that $\mce_F > {\mceo}$, i.e., when both spin branches are
filled. Note that this cancelation only occurs in the $B\to0$ limit. Hence,
since $\sigma^{(1)}$ identically vanishes for any $\mce_F$ and $B$, we find
that in the $B\to0$ limit $\sigma^z_{xy}$ is non-vanishing only for ${\mcee} <
\mce_F < {\mceo}$ being given by
\begin{equation}
  \label{eq:sigtotal}
  \sigma^z_{xy}
  = \sigma^{(2)}
  = \sigma_0
  \left[
    \frac{1}{\kappa_1}\left(\frac{1}{2\kappa_3} - 1\right)
    +
    \frac{\kappa_2 + \kappa_1/2}{4\kappa_3^3}
  \right].
\end{equation}
In the above expressions, the Fermi energy $\mce_F$ (at $B=0$) is given by
\begin{align}
  \mce_{F}
  & =
  \begin{cases}
    \displaystyle
    \frac{\pi\hbar^{2}n_{2D}}{2m}
    + \bar{\mce} + \frac{m\eta^2}{\hbar^2}\,,
    & \mce_F {>} \mce_{o}
    \\
    \displaystyle
    \frac{\pi\hbar^{2}n_{2D}}%
    {m\left(1+\dfrac{2m\eta^2}{\hbar^2 \Delta\mce}\right)}
    + \mce_{e}\,,
    & \mce_{e} {<} \mce_F {<} \mce_{o}\,
  \end{cases}
  \label{eq:efformula}
\end{align}
as can be straightforwardly derived from the $B=0$ spectrum of our
system.\cite{expansion} Next we calculate $\sigma^z_{xy}$ for realistic wells.

\section{Self consistent calculation and results}

\begin{figure}[!b]
  \centering
  \resizebox{\imsizeB}{!}{\includegraphics{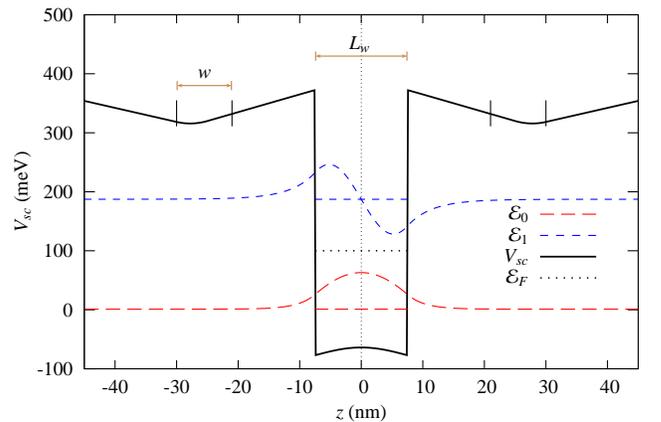}}
  \caption{Self consistent potential profile of our InSb/Al$_{0.3}$In$_{0.7}$Sb
    quantum well with two subbands. The calculated subband edges
    $\mce_{e}=1.0$~meV and $\mce_{o}=187.3$~meV and the corresponding
    wavefuntions are also shown. The Fermi energy is $\mce_{F}=100.3$~meV.}
  \label{fig:sqw}
\end{figure}

We have performed a detailed self-consistent calculation -- by solving the
Schr{\"{o}}dinger and Poisson's equations in the Hartree approximation -- to
determine the intersubband SO coupling strength $\eta$ for realistic symmetric
wells\cite{Calsaverini08} and the corresponding spin Hall conductivity
$\sigma_{xy}^{z}(\eta)$. We have considered a modulation-doped symmetric
Al$_{0.3}$In$_{0.7}$Sb/InSb/Al$_{0.3}$In$_{0.7}$Sb quantum well with only two
subbands;\cite{parameters} see \figref{fig:sqw}. The structure comprises a
single well of width $L_{w}=15$~nm and two n-doped semi-infinite adjacent
regions. We assume that the ionized impurities giving up electrons to the well
form a continuum positive background with density $\rho$ of width $w=5$ nm
(depletion layer). Our self consistent calculation follows that of
Ref.~[\onlinecite{Calsaverini08}].

\begin{figure}[!t]
  \centering
  \resizebox{\imsizeB}{!}{\includegraphics{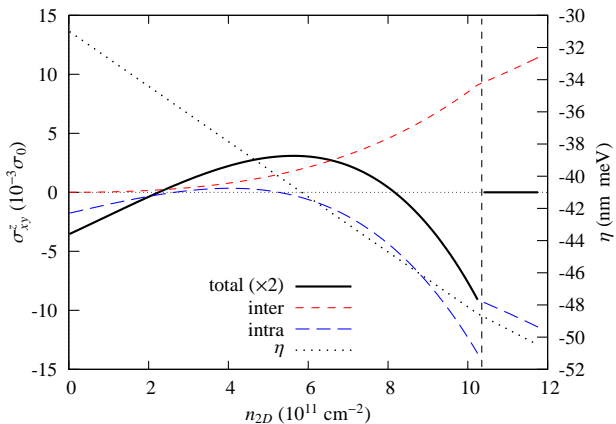}}
  \caption{Interbranch (dashed line) and intrabranch (long-dashed line)
    conductivity contributions and the total (solid line) spin Hall
    conductivity $\sigma^z_{xy}$ (in units of $\sigma_0=e/4\pi$) as functions
    of the areal density $n_{2D}$ in the single InSb well shown in
    \figref{fig:sqw}.  Note the discontinuities at $\mce_{F}=\mce_{o}$ and
    $\mce_F=\mce_e$, which correspond to $n_{2D}=10.3\times 10^{11}$~cm$^{-2}$
    and $n_{2D}=0$, respectively.}
  \label{fig:shc2}
\end{figure}

In an earlier work,\cite{Calsaverini08} we have found that narrow single InSb
wells give sizable values for the intersubband-induced SO coupling strength
$\eta$. Here we self consistently calculate the corresponding spin Hall
conductivity as a function of the areal density $n_{2D}$.\cite{chem-pot}
\Figref{fig:shc2} shows the calculated interbranch ($\sigma_{\rm
  inter}=\sigma^{(1)}_{\rm inter} + \sigma^{(2)}_{\rm inter}$) and intrabranch
($\sigma_{\rm intra}=\sigma^{(1)}_{\rm intra} + \sigma^{(2)}_{\rm intra}$)
conductivity contributions and the total spin Hall conductivity $\sigma^z_{xy}$
[see \eqnref{eq:sigtotal}] as a function of the areal density $n_{2D}$ (or
$\mce_F)$ for a single InSb well.  Note the two discontinuities of the SHC at
densities corresponding to $\mce_F = {\mceo}$ (see vertical dashed line) and
$\mce_F=\mce_{e}$ (at $n_{2D}=0$).  The magnitude of these jumps are
$\Delta\sigma/\sigma_0 = \kappa_1/(1 + \kappa_1)^2$ and $-\kappa_1/(1 -
\kappa_1)^2$, respectively. These discontinuities come from the intrabranch
contribution $\sigma^z_{xy, \rm intra} = \sigma^{(2)}_{\rm intra}$, while the
interbranch contribution $\sigma^z_{xy,\rm inter} = \sigma^{(2)}_{\rm inter}$
varies smoothly with $\mce_F$ (recall that $\sigma^{(1)}=0$). In addition, the
competition between $\sigma^{(2)}_{\rm intra}$ and $\sigma^{(2)}_{\rm inter}$
can lead to a sign change of $\sigma^z_{xy}$ as $\mce_F$ is varied, see
\figref{fig:shc2}.  It should be noted that as the strength of the SO coupling
$\eta$ (or $\kappa_1$) becomes smaller, $\Delta\sigma$ and $\sigma^z_{xy}$
diminish as well, thus vanishing completely at $\kappa_1 = 0$.

Despite the interesting features displayed by the total spin Hall conductivity
-- the sign change and the non-monotonic behavior -- as the electron density is
varied between the subband edges, we find that the largest values of $
\sigma^z_{xy}$ are extremely small ($\sim 10^{-3} \sigma_0$).

\section{Additional discussion}

Considering the formal similarity between our Hamiltonian containing the
intersubband-induced SO coupling and Rashba's, it is not entirely surprising
that the SHC vanishes when both spin branches are filled. Novel effects of the
intersubband-induced SO term arise when only the lower branch is filled: an
incomplete cancelation between the contributions from the interbranch and
intrabranch transitions lead to a nonzero SHC and discontinuities in it. The
origin of these discontinuities can be traced back to the abrupt changes in
$B=0$ density of states (DOS) of the model at the subband edges. The $B=0$ DOS
for each spin branch is given by $\rho_\mp = \frac12 \rho_0 (1 \mp
\kappa_1/2\kappa_3)$ with $\rho_0=m/\pi\hbar^{2}$. Hence the DOS abruptly
changes at $\mce_F = \mce_{\rm o}$ from a constant value $\rho_+ + \rho_- =
\rho_0$ to $\rho_+$, thus giving rise to an abrupt loss of states contributing
to the spin density response.  In fact, the intrabranch contribution that is
responsible for the discontinuity is proportional to the DOS, that is,
$1/(\omega_{\lambda\mu n_{\lambda\mu}} - \omega_{\lambda\mu
  n_{\lambda\mu}{+}1}) \propto \rho_\mu$ [see \eqnref{eq:shc2}]. The
discontinuity $\Delta\sigma$ in $\sigma_{xy}^{z}$ can be explicitly related to
the discontinuity $\Delta\rho$ in the DOS as follows
\begin{align}
  \frac{\Delta\sigma}{\sigma_0}
  & =
  \begin{cases}
    \displaystyle
    2\frac{\Delta\rho}{\rho_0}\left(1-2\frac{\Delta\rho}{\rho_0}\right)\,,
    & \mce_F {=} \mce_{\rm o}
    \\
    \displaystyle
    -2\frac{\Delta\rho}{\rho_0}\left(1+2\frac{\Delta\rho}{\rho_0}\right)\,,
    &  \mce_F {=} \mce_{\rm e}\,.
  \end{cases}
\end{align}
Note that both expressions vanish as $\Delta\rho/\rho_0 \rightarrow 0$, clearly
showing that $\Delta\sigma$ arises from the discontinuity in the DOS.

Our intersubband-induced SO coupling mixes both the spin and the subband
degrees of freedom simultaneously. Therefore it does not provide a mechanism to
couple \textit{opposite} spins within a given subband. This implies that any
projection of our Hamiltonian into the lower subband, in the limit of large
subband gap $\Delta\mce \gg \mces$, would not produce an effective SO coupling
between spins within the lower subband so that no finite SHC appears.  This
limiting case study stresses that our finite SHC, even for Fermi energies near
the lower subband bottom, is due to the subband-transfer process, even though
it should be quite small in the limit $\Delta\mce \gg \mces$. We expect,
however, that additional subband-mixing but spin-preserving scattering
mechanisms, such as impurity scattering, can induce an effective SO coupling
within a given subband when mediated by the intersubband-induced SO coupling,
which mixes both spins and subbands. This impurity-mediated SO coupling may
then affect and possibly even enhance the strength of the SHC, as long as its
momentum randomization is weak enough.

As a final point we mention that interbranch contribution $\sigma^z_{xy,\rm
  inter}$ reproduces the result for the SHC calculated via the Kubo formula in
the absence of a magnetic field.  Note that this quantity is non universal and
vanishes as $\eta\to0$. Nevertheless, we stress that it alone does not
constitute the total SHC.

\section{Summary and final remarks}

We have calculated the ballistic spin Hall conductivity $\sigma^z_{xy}$ for
symmetric wells with two subbands in which the intersubband-induced SO
interaction is present. We follow a linear response approach due to Rashba
which consistently accounts for intrabranch and interbranch transitions in the
Kubo formula. We find that $\sigma^z_{xy}$ is zero when the two subbands are
occupied (similar to the Rashba model) and non-zero when only the lower subband
is occupied. We have also performed a numerical self consistent calculation to
determine the intersubband SO strength for realistic InSb wells and have
calculated the corresponding $\sigma^z_{xy}$.  Even though the calculated
$\sigma^z_{xy}$ shows interesting features such as discontinuities at the
subband edges (due to discontinuities in the DOS), a non-monotonic behavior and
a sign change as a function of the Fermi energy (or areal density), the
magnitude of $\sigma^z_{xy}$ is much smaller than 1 in units of $e/4\pi$.

It is conceivable that other materials systems, e.g. metallic surfaces and
interfaces, can display intersubband induced SO interaction, in addition to the
usual Rashba. For instance, metallic surfaces with unconventional spin
topology\cite{Mirhosseini09} where deviations from the usual Rashba model have
recently been reported (and for which the SO couplings are much stronger than
in semiconductors) are a possibility. Perhaps in these systems the SHC due to
the intersubband SO coupling would be sizable.

One caveat of our calculation is that we use a spin current definition which is
a simple extension for two subbands of the conventional (not-uniquely defined
\cite{sugi}) symmetrized product of the spin and velocity operators as in
Rashba model. Hence all the issues related to the reality of these currents and
whether or not they would lead to spin accumulation in finite samples appear
here as well. More work is certainly needed to address these
issues. \cite{mish} Using non-equilibrium Green functions on a lattice, we have
performed some simulations\cite{siggi} of the spin density in bilayer systems
with inter-layer SO orbit coupling, whose Hamiltonian maps onto our two-subband
one. Our preliminary results show that the spin density changes as compared to
the single layer case (Rashba model). Finally, we emphasize that the role of
impurities, which we believe should not kill the effect discussed here, remains
an interesting problem for further investigations.

\begin{acknowledgments}
  JCE acknowledges useful discussions with John Schliemann.  This work was
  supported by the Swiss NSF, the NCCR Nanoscience, JST ICORP, CNPq and FAPESP.
\end{acknowledgments}


\begin{thebibliography}{99}

\bibitem{Engel07}
  For reviews, see H.-A. Engel, E. I. Rashba, and B. I. Halperin,
  \textit{Handbook of Magnetism and Advanced Magnetic Materials} (John Wiley \& Sons Ltd, Chichester, 2007)
  and J. Schliemann, Int. J. Mod. Phys. B \textbf{20}, 1015 (2006),

\bibitem{ESHE}
  M. I. Dyakonov and V. I. Perel, Zh. Eksp. Ter. Fiz. Pis'ma Red. {\bf 13}, 657 (1971) [JETP Lett.\textbf{13}, 467 (1971)];
  J. E. Hirsch, Phys. Rev. Lett. \textbf{83}, 1834 (1999);
  S. Zhang, Phys. Rev. Lett. \textbf{85}, 393 (2000);
  L. Hu, J. Gao, and S.-Q. Shen, Phys. Rev. B \textbf{68}, 115302 (2003);
  M. I. Dyakonov, Phys. Rev. Lett. \textbf{99}, 126601 (2007).

\bibitem{Murakami03}
  S. Murakami, N. Nagaosa, and S. C. Zhang, Science \textbf{301}, 1348 (2003);
  Phys. Rev. B \textbf{69}, 235206 (2004).

\bibitem{Sinova04}
  J. Sinova, D. Culcer, Q. Niu, N. A. Sinitsyn, T. Jungwirth, and A. H. MacDonald, Phys. Rev. Lett. \textbf{92}, 126603 (2004).

\bibitem{Inoue04}
  J. I. Inoue, G. E. W. Bauer, and L. W. Molenkamp, Phys. Rev. B \textbf{70}, 041303(R) (2004).

\bibitem{Mishchenko04}
  E. G. Mishchenko, A. V. Shytov, and B. I. Halperin, Phys. Rev. Lett. \textbf{93}, 226602 (2004).

\bibitem{raimondi05}
  R. Raimondi and P. Schwab, Phys. Rev. B \textbf{71}, 033311 (2005)

\bibitem{Chalaev05}
  O. Chalaev and D. Loss, Phys. Rev. B \textbf{71}, 245318 (2005).

\bibitem{Dimitrova05}
  O. V. Dimitrova, Phys. Rev. B \textbf{71}, 245327 (2005).

\bibitem{Malshukov05}
  A. G. Mal'shukov and K. A. Chao, Phys. Rev. B \textbf{71}, 121308(R) (2005).

\bibitem{Shen}
  S.-Q. Shen, M. Ma, X. C. Xie, and F. C. Zhang, Phys. Rev. Lett. \textbf{92}, 256603 (2004);
  S.-Q. Shen, Y.-J. Bao, M. Ma, X. C. Xie, and F. C. Zhang, Phys. Rev. B \textbf{71}, 155316 (2005).

\bibitem{Gorini08}
  C. Gorini, P. Schwab, M. Dzierzawa, and R. Raimondi, Phys. Rev. B \textbf{78}, 125327 (2008).

\bibitem{Rashba04}
  E. I. Rashba, Phys. Rev. B \textbf{70}, 201309(R) (2004).

\bibitem{Lucignano08}
  P. Lucignano, R. Raimondi, and A. Tagliacozzo, Phys. Rev. B \textbf{78}, 035336 (2008).

\bibitem{Kato04}
  Y. K. Kato, R. C. Myers, A. C. Gossard, and D. D. Awschalom, Science \textbf{306}, 1910 (2004);
  N. P. Stern, D. W. Steuerman, S. Mack, A. C. Gossard, and D. D. Awschalom, Nat. Phys. \textbf{4}, 843 (2008).

\bibitem{Wunderlich05}
  J. Wunderlich, B. Kaestner, J. Sinova, and T. Jungwirth, Phys. Rev. Lett. \textbf{94}, 047204 (2005).

\bibitem{Schliemann05}
  J. Schliemann and D. Loss, Phys. Rev. B \textbf{71}, 085308 (2005).

\bibitem{Sih05}
  V. Sih, R. C. Myers, Y. K. Kato, W. H. Lau, A. C. Gossard, and D. D. Awschalom, Nat. Phys. \textbf{1}, 31 (2005).

\bibitem{Sinova06}
  J. Sinova, S. Murakami, S.-Q. Shen, and M.-S. Choi, Solid State Communications \textbf{138}, 214 (2006).

\bibitem{Erlingsson05}
  S. I. Erlingsson, J. Schliemann, and D. Loss, Phys. Rev. B \textbf{71}, 035319 (2005).

\bibitem{non-zero-she} We note, however, that finite size effects and quantum
  interference can still lead to coherent spin accumulation or non-trivial spin
  conductance, provided that the sample size is comparable to the spin
  precession length.\cite{Nikolic,Lee05,Reynoso05,Yao06} Moreover, for
  oscillating driving fields\cite{Duckheim06} ESR type phenomenon is possible in
  Rashba coupled 2DEGs in the presence of an in-plane magnetic field and a
  non-zero SHC emerges in the high frequency (as compared to the inverse
  momentum relaxation time) limit. We should also mention that $p$-doped wells
  can exhibit a considerable intrinsic SHE\cite{Schliemann05} since the SO
  interaction for heavy holes is cubic in the carrier momentum. For instance,
  the electrical detection of the intrinsic SHE in HgTe quantum wells via
  non-local resistance measurements shows an enhanced signal as the gate
  voltage controlling the carrier density is tuned from $n$-type to $p$-type
  conduction.\cite{Brune08}

\bibitem{Koenig09}
  M. Koenig, H. Buhmann, L. W. Molenkamp, T. L. Hughes, C.-X. Liu, X.-L. Qi, and S.-C. Zhang, arXiv:0801.0901, special issue of J. Phys. Soc. Jpn. (to be published);
  S. Takahashi and S. Maekawa, Sci. Technol. Adv. Mater. \textbf{9}, 014105 (2009).

\bibitem{Bernardes06}
  E. Bernardes, J. Schliemann, M. Lee, J. C. Egues, and D. Loss, Phys. Rev. Lett. \textbf{99}, 076603 (2007).

\bibitem{Calsaverini08}
  R. S. Calsaverini, E. Bernardes, J. C. Egues, and D. Loss, Phys. Rev. B \textbf{78}, 155313 (2008).

\bibitem{spin-qc} This is ``the quantum of spin conductance'', the analog to
  the quantum of charge conductance $g=e^2/h$; these are related by
  $\sigma_0=(\hbar/2e)g=e/4\pi$.

\bibitem{expansion} For the $B=0$ the energy spectrum is
  $\mce_{k_\|\lambda\sigma}=\hbar^2k_{\|}^2/2m + \bar\mce + \lambda
  \sqrt{(\Delta\mce/2)^2+\eta^2 k_\|^2}$, where $k_\|$ is the in-plane electron
  wave vector, $\lambda=\pm$ the subband index and $\sigma=\uparrow,
  \downarrow$ the spin index. To obtain the expression, \eqnref{eq:efformula}
  for $\mce_F$, we assume that $\eta k_\|<< \Delta\mce$ (this is a good
  approximation for the areal densities used).

\bibitem{parameters} All other relevant band parameters can be found in
  Table~III of Ref.~[\onlinecite{Calsaverini08}] (see also
  Ref.~[\onlinecite{Vurgaftman01}] for further information on the band
  structure parameters).

\bibitem{chem-pot} Similarly to Ref.~[\onlinecite{Calsaverini08}], here we
  perform the self consistent calculation to determine the intersubband SO
  couplint $\eta$ by considering a quantum well with a constant chemical
  potential (see Refs.~[\onlinecite{Koga02}] and [\onlinecite{Koga06}] for wells
  with constant chemical potentials); as in Ref.~[\onlinecite{Calsaverini08}],
  similar results can be obtained for the case of a constant density. The areal
  density $n_{2D}$ of the well is varied via an additional gate which shifts
  the bottom of the potential well with respect to the fixed chemical potential
  level.

\bibitem{Mirhosseini09}
  H. Mirhosseini, J. Henk, A. Ernst, S. Ostanin, C.-T. Chiang, P. Yu, and A. Winkelmann, and J. Kirschner, Phys. Rev. B \textbf{79}, 245428 (2009).

\bibitem{sugi}
  N. Sugimoto, S. Onoda, S. Murakami, and N. Nagaosa, Phys. Rev. B \textbf{73},113305 (2006).

\bibitem{mish}
  P. G. Silvestrov, V. A. Zyuzin, and E. G. Mishchenko, Phys. Rev. Lett. \textbf{102}, 196802 (2009).

\bibitem{siggi}
  S. I. Erlingsson, J. Carlos Egues, and Daniel Loss, Physica E: Low-dimensional Systems and Nanostructures \textbf{40}, 1484 (2008).

\bibitem{Nikolic}
  B. K. Nikoli\'c, S. Souma, L. P. Z\^arbo, and J. Sinova, Phys. Rev. Lett. \textbf{95}, 046601 (2005);
  B. K. Nikoli\'c, L. P. Z\^arbo, and S. Souma, Phys. Rev. B \textbf{72}, 075361 (2005);
  S. Souma and B. K. Nikoli\'c, Phys. Rev. Lett. \textbf{94}, 106602 (2005);
  G. Usaj and C. Balseiro, Europhys. Lett. \textbf{72}, 631 (2005).

\bibitem{Lee05}
  M. Lee and M.-S. Choi, Phys. Rev. B \textbf{71}, 153306 (2005).

\bibitem{Reynoso05}
  A. Reynoso, G. Usaj, and C. A. Balseiro, Phys. Rev. B \textbf{73}, 115342 (2006)

\bibitem{Yao06}
  J. Yao and Z. Q. Yang, Phys. Rev. B \textbf{73}, 033314 (2006).

\bibitem{Duckheim06}
  M. Duckheim and D. Loss, Nat. Phys. \textbf{2}, 195 (2006).

\bibitem{Brune08}
  C. Br\"une, A. Roth, E. G. Novik, M. K\"onig1 H. Buhmann, E. M. Hankiewicz,
  W. Hanke, J. Sinova, and L. W. Molenkamp, arXiv:0812.3768 (unpublished).

\bibitem{Vurgaftman01}
  I. Vurgaftman, J. R. Meyer, and L. R. Ram-Mohan, J. of Appl. Phys. \textbf{89}, 5815 (2001).

\bibitem{Koga02}
  T. Koga, J. Nitta, T. Akazaki, and H. Takayanagi, Phys. Rev. Lett. \textbf{89}, 046801 (2002).

\bibitem{Koga06}
  T. Koga, Y. Sekine, and J. Nitta, Phys. Rev. B \textbf{74}, 041302(R) (2006).
\end{thebibliography}
\end{document}